\documentclass[runningheads]{svmult}

\usepackage{makeidx}   
\usepackage{graphicx}  
\usepackage{subeqnar}  
\usepackage{multicol}  
\usepackage{physprbb}  
\makeindex             

\newcommand{\gtsim}{\mbox{{\raisebox{-0.4ex}{$\stackrel{>}{{\scriptstyle\sim}}
$}}}}
\newcommand{\ltsim}{\mbox{{\raisebox{-0.4ex}{$\stackrel{<}{{\scriptstyle\sim}}
$}}}}

\begin{document}
\title*{The mass of radio galaxies from low to high redshift}
\toctitle{The mass of radio galaxies from low to high redshift}
%
%
\titlerunning{The mass of radio galaxies}
%
\author{Matt J. Jarvis\inst{1}
\and Steve Rawlings\inst{2}
\and Steve Eales\inst{3}
\and Katherine M. Blundell\inst{2}
\and Chris J. Willott\inst{2}}
\authorrunning{M. J. Jarvis et al.}
%
%
\institute{Sterrewacht Leiden, Postbus 9513, 2300 RA Leiden, The Netherlands
\and Astrophysics, Department of Physics, Keble Road, Oxford, OX1 3RH, UK
\and Department of Physics and Astronomy, University of Wales College of Cardiff, P.O.Box 913, Cardiff, CF2 3YB, UK}

\maketitle              

\begin{abstract}
Using a new radio sample, 6C* designed to find radio galaxies at $z >
 4$ along with the complete 3CRR and 6CE sample we extend the radio
 galaxy $K-z$ relation to $z \sim 4.5$.  The 6C* $K-z$ data
 significantly improve delineation of the $K-z$ relation for radio
 galaxies at high redshift ($z >2$). In a spatially flat universe with
 a cosmological constant ($\rm \Omega_{\rm M}=0.3$ and
 $\rm \Omega_{\Lambda}=0.7$), the most luminous radio sources appear to be
 associated with galaxies with a luminosity distribution with a high
 mean ($\approx 5L^{\star}$), and a low dispersion ($\sigma \sim
 0.5$~mag) which formed their stars at epochs corresponding to $z\,
 \gtsim\, 2.5$.
\end{abstract}

\section{The advantages of using radio galaxies}

Radio galaxies provide the most direct method of investigating the
host galaxies of quasars if orientation based unified schemes are
correct. The nuclear light which dominates the optical/near-infrared
emission in quasars is obscured by the dusty torus in radio galaxies,
therefore difficult psf modelling and subtraction
are not required to determine the properties of the underlying host
galaxy. Unfortunately compiling samples of radio loud AGN is a long
process, because of the radio selection there is no intrinsic optical
magnitude limitation, making follow-up observations extremely time
consuming, especially when dealing with the faintest of these objects.
However, low-frequency selected radio samples do now exist with the
completion of 3CRR (Laing, Riley \& Longair 1983) along with 6CE
(Eales et al. 1997; Rawlings et al. 2001) and the filtered 6C* sample
(Blundell et al. 1998; Jarvis et al. 2001a; 2001b). We can now use
these radio samples to investigate the underlying stellar populations
through the radio galaxy $K-z$ Hubble diagram.

\section{Previous radio samples and the $K-z$ Hubble diagram}
There has been much interest in the $K-z$ relation for radio galaxies
in the past decade. Dunlop \& Peacock (1993) using radio galaxies from
the 3CRR sample along with fainter radio sources from the Parkes
selected regions demonstrated that there exists a correlation between
radio luminosity and the $K-$band emission. Whether this is due to a
radio luminosity dependent contribution from a non-stellar source or
because the galaxies hosting the most powerful radio sources are
indeed more massive galaxies has yet to be resolved. Eales et
al. (1997) confirmed this result and also found that the dispersion in
the $K-$band magnitude from the fitted straight line increases with
redshift. This result, along with the departure to brighter magnitudes
of the sources at high redshift led Eales et al. to conclude that we
are beginning to probe the epoch of formation of these massive
galaxies. Using the highest redshift radio galaxies from ultra-steep
samples of radio sources van Breugel et al. (1998) found that the near
infrared colours of radio galaxies at $z > 3$ are very blue,
consistent with young stellar populations. They also suggest that the
size of the radio structure is comparable with the size of the near
infrared region, and the alignment of this region with the radio
structure is also more pronounced at $z > 3$. Lacy et al. (2000) using
the 7C-III sample found evidence that the hosts of radio galaxies
become more luminous with redshift and are consistent with a passively
evolving population which formed at high redshift ($z > 3$). Thus, all
of this work points to a radio galaxy population which formed at high
redshift and has undergone simple passive evolution since. However,
all of these studies were made with only a few high-redshift ($z > 2$)
sources. With the 6C* sample we are now able to probe this
high-redshift regime with increased numbers from samples with
well-defined selection criteria.

\section{The 6C* filtered sample}
\begin{figure}[t]
\begin{center}
\includegraphics[width=1.\textwidth]{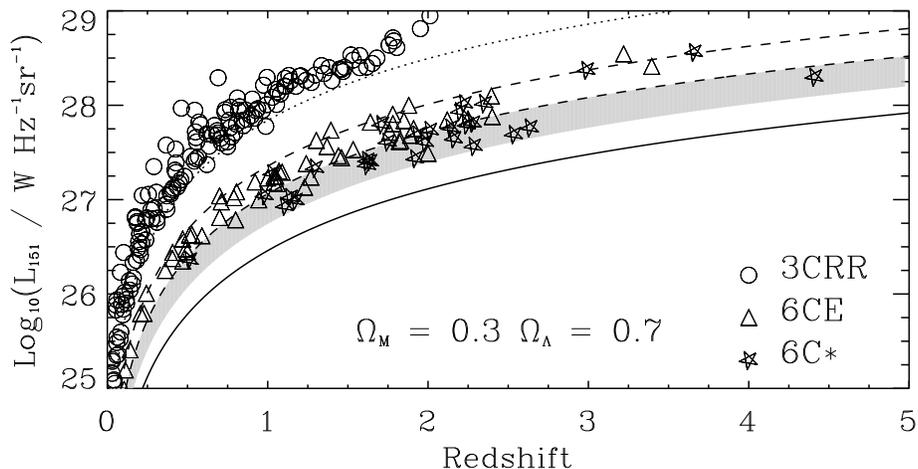}
\end{center}
\caption[]{Rest-frame
151\,MHz luminosity ($L_{151}$) versus redshift $z$ plane for the 3CRR
(circles), 6CE (triangles) and 6C* (stars) samples.  The rest-frame
151\,MHz luminosity $L_{151}$ has been calculated according to a
polynomial fit to the radio spectrum (relevant radio data from
Blundell et al. 1998).  The curved lines show the lower flux-density
limit for the 3CRR sample (dotted line; Laing et al. 1983) and the
7CRS (solid line; Blundell et al. in prep; Willott et al. in
prep). The dashed lines correspond to the limits for the 6CE sample
(Rawlings et al. 2001) and the shaded region shows the 6C*
flux-density limits (all assuming a low-frequency radio spectral index
of 0.5). Note that the area between the 3CRR sources and 6CE sources
contains no sources, this is the area which corresponds to the absence
of a flux-density limited sample between the 6CE ($S_{151} \leq
3.93\:$Jy) and 3CRR ($S_{178} \geq 10.9\:$Jy) samples. The reason why
some of the sources lie very close to or below the flux-density limit
of the samples represented by the curved lines is because the spectral
indices lie very close to or below the assumed spectral index of the
curves of $\alpha = 0.5$.}
\label{fig:pzplane}
\end{figure}

The 6C* sample is a low-frequency radio sample ($0.96 {\rm Jy} \leq
S_{151} \leq 2.00$\,Jy) which was originally designed to find radio
sources at $z>4$ using filtering criteria based on the radio
properties of steep spectral index and small angular size. The
discovery of 6C*0140+326 at $z = 4.41$ (Rawlings et al. 1996) and
6C*0032+412 at $z = 3.66$ (Jarvis et al. 2001a) from a sample of just
29 objects showed that this filtering was indeed effective in finding
high-redshift objects. Indeed, the median redshift of the 6C* sample
is $z \sim 1.9$ whereas for complete samples at similar flux-density
levels the median redshift is $z \sim 1.1$. We can now use this sample
to push the radio galaxy $K-z$ diagram to high redshift ($z > 2$)
where it has not yet been probed with any significant number of
sources. Fig.~\ref{fig:pzplane} shows the radio luminosity-redshift
plane for the three samples used in this analysis.

\section{Emission-line contamination}
The most-luminous sources at high redshift may be contaminated by the
bright optical emission lines redshifted into the infrared. This is
particularly true for sources in radio flux-density limited
samples. The high-redshift sources in these samples are inevitably
some of the most luminous, and we also know there is a strong
correlation between low-frequency radio luminosity and emission-line
strength (e.g. Rawlings \& Saunders 1991; Willott et al. 1999; Jarvis
et al. 2001a) which will increase the contribution to the measured
$K-$band magnitudes from the emission lines in the most radio luminous
sources.

\begin{figure}[t]
\begin{center}
\includegraphics[width=1.\textwidth]{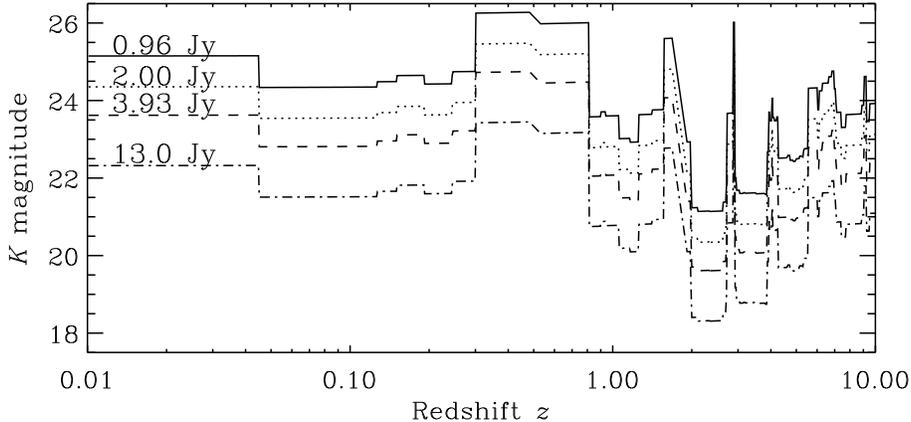}
\end{center}
\caption[]{Emission line contribution to the
$K-$band magnitudes for various radio flux-densities assuming the
power-law relation of $L_{\rm [OII]} \propto L_{151}^{1.00}$.}
\label{fig:line_cont}
\end{figure}

To subtract this contribution we use the correlation between [OII]
emission-line luminosity $L_{\rm [OII]}$ and the low-frequency radio
luminosity $L_{151}$ from Willott (2000), where $L_{\rm [OII]} \propto
L_{151}^{1.00 \pm 0.04}$. Then by using the emission-line flux ratios
for radio galaxies (e.g. McCarthy 1993) we are able to determine the
contribution to the $K-$band magnitude from all of the other
emission lines. This is illustrated in Fig.~\ref{fig:line_cont} where
the emission-line contamination to the $K-$band flux is shown for
various radio flux-density limits and a range of redshifts.

\section{The $K-z$ relation}
\begin{figure}[t]
\begin{center}
\includegraphics[width=0.8\textwidth]{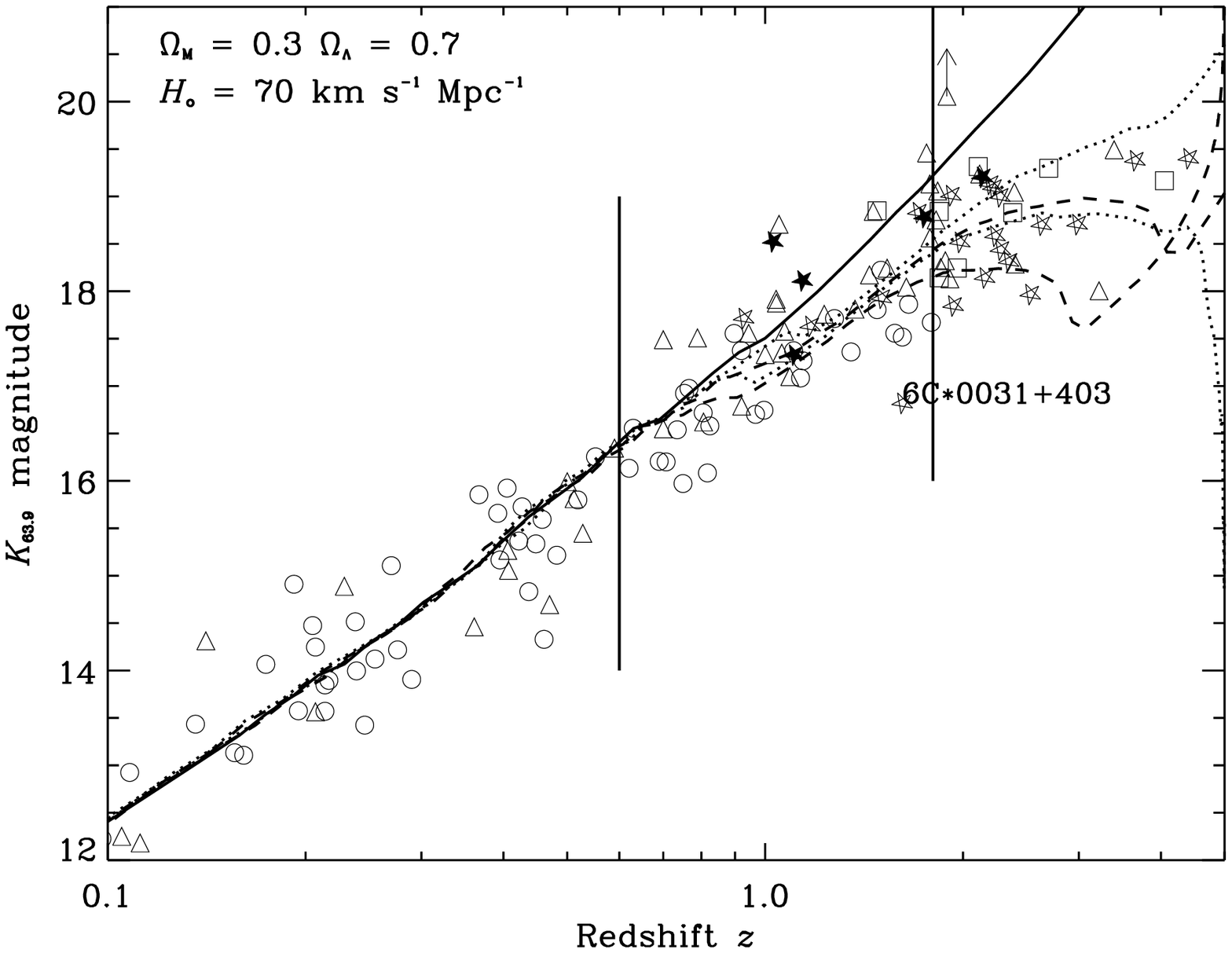}
\end{center}
\caption[]{The $K-z$ Hubble diagram for radio galaxies
from the 3CRR (circles), 6CE (triangles), 6C* (stars) and 7C-III
(squares) samples. $K_{63.9}$ denotes the $K-$band magnitude within a
comoving metric aperture of 63.9 kpc (c.f. Eales et al. 1997; Jarvis
et al. 2001b).  The two vertical lines show the redshift above which
the alignment effect begins to be seen ($z > 0.6$) and the higher
redshift at which we chose to split the data ($z = 1.8$). The solid
curved line is the predicted curve for galaxies which do not evolve
(as described in the text). The dashed lines are models for a
star-burst lasting 1 Gyr starting at $z = 5$ (lower) and $z = 10$ (upper). The two
dotted curves represent the models of an instantaneous (0.1 Gyr) burst
beginning at $z =5$ (lower) and $z = 10$ (upper). }
\label{fig:fig:kzs}
\end{figure}

In Fig.~3 we plot the $K-z$
diagram for all of the sources in our dataset. We also show four
synthetic galaxy evolution models from the `Galaxy Isochrone Synthesis
Spectral Evolution Library' (GISSEL) of Bruzual \& Charlot (1993) and
a curve representing a galaxy which undergoes no-evolution. The GISSEL files that we have used
are ones in which there is an instantaneous burst of star formation
and one in which the burst of star formation lasts 1 Gyr, with a
Salpeter IMF with a lower mass cut-off of 0.1 M$_{\odot}$ and an upper
mass cut-off of 125 M$_{\odot}$. We use two different assumptions
about the star formation history, one in which the burst of star
formation begins at $z = 5$ and one in which the burst occurs at $z =
10$.

The no-evolution curve was constructed by taking the spectral energy
distribution template from the GISSEL library 
that was found to fit the observed spectral energy distribution of
a radio galaxy at $z = 0$, and which also reproduced the near-infrared
colours. All of the curves are normalised to pass through the
low-redshift ($z < 0.3$) points.

With our data on the 6C* sample in addition to the 6CE and 7C-III
samples we find that in a low-density universe ($\rm \Omega_{\rm M}=0.3$ and $\rm \Omega_{\Lambda}=0.7$) the data are
predominantly brighter than the no-evolution curve and are consistent
with a passively evolving stellar population with a high-formation
redshift. If this passively evolving scenario is correct then
hierarchical growth at $z < 2.5$ is not a required ingredient.  However,
this brightening may not just be due to passive evolution of the
stellar population. Non-stellar contributions from the central AGN may
also contribute a higher fraction of light at these redshifts. All of
the studies to measure the non-stellar contribution to the $K-$band flux
conducted to date (e.g. Leyshon \& Eales 1998; Simpson, Rawlings \&
Lacy 1999), have concentrated on the most radio luminous 3CR sources
at $z \simeq 1$, and may have little bearing on the properties of the
high-redshift sources considered here. If it turns out that
the high-redshift 6C sources have $\gtsim$\, non-stellar contamination
to those of the $z \sim 1$ 3C sources (which have the same radio
luminosity) then hierarchical build up may be necessary. Note that
$K-$band observations of the high-redshift 6C sources will be at shorter
wavelengths than those of the 3C sources.

However, separate arguments lead us to conclude that the dominant
factor is passive evolution of a stellar population which formed at
$z\, \gtsim \,2.5$. First, recent sub-mm observations with SCUBA have
shown that the dust masses in radio galaxies are larger at $z \simeq
3$ than in galaxies with similar radio luminosities at lower redshift
(Archibald et al. 2001). This implies that the majority of
star-formation activity in these galaxies is occurring at high
redshift. Second, the discovery of six extremely red objects at $1 < z
< 2$ in the 7C Redshift Survey (Willott, Rawlings \& Blundell 2001)
with inferred ages of a few Gyrs, implies that these objects formed
the bulk of their stellar population at $z \simeq 5$. Third, detailed
modelling of the optical spectrum of the weak radio source LBDS 53W091
at $z = 1.552$ has shown that this object is most plausibly an old
elliptical, with an inferred age of $\gtsim 3.5$~Gyr (Dunlop et
al. 1996). The further discovery of LBDS 53W069 at $z = 1.43$, with an
inferred age of $\sim 4$~Gyr (Dunlop 1999) suggests that there exists
a population of evolved, radio weak ellipticals which formed at $z
\gtsim 5$. Therefore, the new data on the 6C* sample presented in
here is consistent with the results from various other
observational studies of radio galaxies in which these radio-luminous
systems formed most of their stars at epochs corresponding to very
high redshifts ($z\, \gtsim\, 2.5$), and have undergone simple passive
stellar evolution since. Willott et al. (2001) have pointed out that
such galaxies probably undergo at least two active phases: one, at
epochs corresponding to $z \gtsim 5$, when the black hole and stellar
spheroid formed, and another, at e.g. $z \sim 2$, when powerful jet
activity is triggered, or perhaps re-triggered, by an event such as an
interaction or a merger. The small scatter in the $K-z$ relation (Jarvis et al. 2001b) and sub-mm results (Archibald et al. 2001) suggest that the
second active phase has little influence on the stellar mass of the
final elliptical galaxy

We have shown that the powerful radio galaxies in our samples are
consistent with having passively evolving stellar populations.  If we
now compare the masses of these powerful radio galaxies to the derived
value of $M_{K}^{\star}$ for nearby elliptical galaxies
[$M_{K}^{\star} = -24.3$ for $H_{\circ} = 70\,{\rm km\,s}^{-1}\,{\rm
Mpc}^{-1}$ (Kochanek et al. 2000)], we find, if passive evolution is
accounted for, that the powerful radio galaxies considered in this
here are consistent with being $\approx 5 L^{\star}$ throughout the
redshift range $0 < z \ltsim\, 2.5$.

%


\begin{thebibliography}{8.}
\addcontentsline{toc}{section}{References}
\bibitem{arch01}
Archibald E.N., et al., 2001, MNRAS, 323, 417

\bibitem{blun98}
Blundell K.M., et al., 1998, MNRAS, 295, 265

\bibitem{bc93}
Bruzual G., Charlot S., 1993, ApJ, 405, 438

\bibitem{dun99}
 Dunlop J.S., 1999, in `The most distant radio galaxies' KNAW Colloquium, Amsterdam, October 1997, eds R\"ottgering et al., Kluwer

\bibitem{DP93}
Dunlop J.S., Peacock J.A., 1993, MNRAS, 263, 936

\bibitem{dun96}
Dunlop J.S., Peacock J.A., Spinrad H., Dey A., Jimenez
R., Stern D., Windhorst R.A., 1996, Nature, 381, 581

\bibitem{Eea97}
Eales S.A., Rawlings S., Law-Green D., Cotter G., Lacy M., 1997, MNRAS, 291, 593

\bibitem{Jea01a}
Jarvis M.J., et al., 2001a, MNRAS, 326, 1563

\bibitem{Jea01b}
Jarvis M.J., et al., 2001b, MNRAS, 326, 1585

\bibitem{koch00}
Kochanek C.S., et al., 2000, ApJ, 543, 131

\bibitem{lbr00}
Lacy M., Bunker A.J., Ridgway S.E., 2000, AJ, 120, 68

\bibitem{lrl}
Laing R.A., Riley J.M., Longair M.S., 1983, MNRAS, 204, 151

\bibitem{le98}
Leyshon G., Eales S.A., 1998, MNRAS, 295, 10

\bibitem{mcc93}
McCarthy P.J., 1993, ARAA, 31, 639

\bibitem{rel01}
Rawlings S., Eales S.A., Lacy M., 2001, MNRAS, 322, 523

\bibitem{raw96}
Rawlings S., et al., 1996, Nature, 383, 502

\bibitem{rs91}
Rawlings S. \& Saunders R., 1991, Nature, 349, 138

\bibitem{simp99}
Simpson C., Rawlings S., Lacy M., 1999, MNRAS, 306, 828

\bibitem{vB98}
van Breugel W.J.M., et al., 1998, ApJ, 502, 614

\bibitem{cjwconf}
Willott C.J., 2000, to appear in Proc. "AGN in their Cosmic Environment", Eds. B. Rocca-Volmerange \& H. Sol, EDPS Conf. Series (astro-ph/0007467)

\bibitem{cjwero}
Willott C.J., Rawlings S., Blundell K.M., 2001, MNRAS, 324, 1

\bibitem{cjwemline}
Willott C.J., Rawlings S., Blundell K.M., Lacy M., 1999, MNRAS, 309, 1017

\end{thebibliography}
\end{document}